\documentclass[preprint,showpacs,preprintnumbers,amsmath,amssymb,superscriptaddress,12pt,graphicx]{revtex4}

\usepackage{graphicx}
\usepackage{dcolumn}
\usepackage{bm}
\usepackage{amssymb}

\begin{document}

\title{Detecting relic gravitational waves in the CMB: Optimal parameters and their constraints}

\author{W.~Zhao}
\email{Wen.Zhao@astro.cf.ac.uk} \affiliation{School of Physics
and Astronomy, Cardiff University, Cardiff, CF24 3AA, United
Kingdom} \affiliation{Wales Institute of Mathematical and
Computational Sciences, Swansea, SA2 8PP, United Kingdom} \affiliation{Department of
Physics, Zhejiang University of Technology, Hangzhou, 310014,
People's Republic of China}
\author{D.~Baskaran}
\email{Deepak.Baskaran@astro.cf.ac.uk} \affiliation{School of
Physics and Astronomy, Cardiff University, Cardiff, CF24 3AA,
United Kingdom} \affiliation{Wales Institute of Mathematical and
Computational Sciences, Swansea, SA2 8PP, United Kingdom}



\small

\begin{abstract}
{\small  The prospect of detecting relic gravitational waves (RGWs), through their imprint in the cosmic microwave background radiation, provides an excellent opportunity to study the very early Universe. In simplest viable theoretical models the RGW background is characterized by two parameters, the tensor-to-scalar ratio $r$ and the tensor spectral index $n_t$. In this paper, we analyze the potential joint constraints on these two parameters, $r$ and $n_t$, using the data from the upcoming cosmic microwave background radiation experiments. Introducing the notion of the best pivot multipole $\ell_t^*$, we find that at this pivot multipole the parameters $r$ and $n_t$ are uncorrelated, and have the smallest variances. We derive the analytical formulae for the best pivot multipole number $\ell_t^*$, and the variances of the parameters  $r$ and $n_t$. We verify these analytical calculations using numerical simulation methods, and find agreement to within $20\%$. The analytical results provides a simple way to estimate the detection ability for the relic gravitational waves by the future observations of the cosmic microwave background radiation.}

\end{abstract}


\maketitle


\section{Introduction \label{s-introduction}}

Detection of relic gravitational waves (RGWs) can be arguably considered one of the most important challenges for current and future cosmic microwave background radiation (CMB) experiments \cite{Planck,WMAP5,groundbasedCMB,balloonbasedCMB,cmbpol}. The RGWs are produced in the early Universe due to the superadiabatic amplification of zero point quantum fluctuations of the gravitational field \cite{a1}. For this reason, the RGWs carry invaluable information about the early history of our Universe that is inaccessible to any other medium (see review \cite{gwreview} for detailed discussion).

A whole range of scenarios of the early Universe, including the inflationary models, generically predict a RGW background with a power-law primordial power spectra \cite{a1,a2a,a2,a3,a4,zhao,others}. In fact, the existence of RGWs is a consequence of quite general assumptions. Essentially, their existence relies only on the validity of general relativity and basic principles of quantum field theory \cite{a1}. The RGW backgrounds are conventionally characterized by two parameters, the so-called tensor-to-scalar ratio $r$ and the primordial power spectral index of RGWs $n_t$ (explained in detail below).

The RGWs leave well understood imprints on the anisotropies in temperature and polarization of CMB \cite{Polnarev,a8,a10,a11,a12,a13}. More specifically, RGWs produce a specific pattern of polarization in the CMB known as the $B$-mode polarization \cite{a8}. Moreover, RGWs produce a negative cross-correlation between the temperature and polarization known as the $TE$-correlation \cite{a12a,a12,polnarev,ours} (see also \cite{Crittenden,Melchiorri}). The theoretical analysis of these imprints along with the data from CMB experiments allows to place constraints on the parameters $r$ and $n_t$ describing the RGW background. 

The current CMB experiments are yet to detect a definite signature of RGWs. It is hoped that, in the near future, with the launch of the Planck satellite \cite{Planck} together with a host of ground-based \cite{groundbasedCMB} and balloon-borne \cite{balloonbasedCMB} CMB experiments as well as the proposed satellite mission CMBPol \cite{cmbpol}, we shall be able to detect a definite signature of the RGW background. In light of this prospect, it is important to be able to effectively constrain the parameters $r$ and $n_t$. A number of papers have discussed the current and potential constraint on the tensor-to-scalar ratio $r$ \cite{currentconstraint}. However, most of these works either ignore the constraint on the spectral index $n_t$, or make simplifying assumptions about its value. One of the common simplifying assumptions is the so-called ``consistency relation" $n_t=-r/8$ \cite{stein,liddle}. It should be noted that, the consistency relation is valid only in the simplest models of inflation namely the single-field slow-roll inflationary model \cite{stein,liddle,polarski}. For a detailed critical discussion of inflationary predictions and data analysis based on these predictions see \cite{Grishchuk0504}. In order to keep our discussion sufficiently general we shall not use this consistency relation in our analysis.

The constraints on the parameters $r$ and $n_t$, characterizing the RGW background, will give us a direct glimpse into the physical conditions in the early Universe. In particular, they will allow to place constraint on the Hubble parameter of the early Universe \cite{GrishchukSolokhin}, which in the case of inflationary models would correspond to the constraints on the energy scale of inflation \cite{stein}. More specifically, the amplitude of the RGW power spectrum at a particular wavelength, characterized by $r$ and $n_t$, determines the Hubble parameter at the time when the particular wavelength left the horizon. Thus, the determination of $r$ and $n_t$ would give a direct measurement of the time evolution of the early Universe, and provide an observational tool to distinguish between the various inflationary type models. In addition, the spectral index $n_t$ has a special character if the RGW background are generated in a primordial Hagedorn phase of string cosmology \cite{q1} or inflation in the loop quantum gravity \cite{q2}, so the determination of $n_t$ provides an observational way to test or rule out these models.  

In this paper we shall analyze the joint constraints on two parameters $r$ and $n_t$ that would be feasible with the analysis of the data from the upcoming CMB experiments. In general, there will be a non-vanishing correlation between parameters $r$ and $n_t$ \cite{cmbpol,colombo}. As will be explained in the following sections, the definition of $r$ and $n_t$ depend on a reference scale characterized by  a multipole number $\ell_0$, which may be chosen arbitrarily. We shall show that with an appropriate choice of this multipole number, which we shall call the best multipole number $\ell^*_t$ (following the terminology of \cite{Knox}), the parameters $r$ and $n_t$ become uncorrelated and have the smallest possible variances. We shall derive approximate analytical expressions for the variances and the correlation coefficients, followed by an analytical calculation of the pivot multipole $\ell_t^*$. Using the Markov Chain Monte Carlo (MCMC) simulation methods, we shall verify our analytical results and evaluate the expected constraints for realistic CMB experiments.

The outline of the paper is as follows. In Section II we shall introduce and explain the notations for the power spectra of gravitational waves, density perturbations and various CMB anisotropy fields and briefly explain how they are calculated. Furthermore, in this section we shall explicitly state the simplifying assumptions that we shall be using throughout the paper, and explain the limits of their applicability. Following this, in Section III, we shall calculate analytically the expected variances and the correlation associated with the parameters $r$ and $n_t$. We shall show existence of the best pivot multipole scale $\ell_t^*$ for which the variances of the corresponding $r$ and $n_t$ are minimal and the correlation between them vanishes. In Section IV we shall confirm our analytical results using numerical calculations. Finally, Section V is dedicated to a brief discussion and conclusions.


\section{Power spectra of cosmological perturbations and CMB fields \label{s-spectra}}

The main contribution to the observed temperature and polarization anisotropies of the CMB comes from two types of the cosmological perturbations, density perturbations (also known as the scalar perturbations) and RGWs (also known as the tensor perturbations) \cite{Polnarev,a3,a4,a8}. These perturbations are generally characterized by their primordial power spectra. These power spectra are usually assumed to be power-law, which is a generic prediction of a wide range of scenarios of the early Universe, including the inflationary models. In general there might be deviations from a power-law, which can be parametrized in terms of the running of the spectral index (see for example \cite{liddle}), but we shall not consider this possibility in the current paper. Thus, the power spectra of the perturbation fields have the form
 \begin{eqnarray}\label{dp-spectrum}
 P_{\mathcal{R}}(k)=A_s(k_0)\left(\frac{k}{k_0}\right)^{n_s-1},
 \end{eqnarray} 
 \begin{eqnarray}\label{gw-spectrum}
 P_{h}(k)=A_t(k_0)\left(\frac{k}{k_0}\right)^{n_t},
 \end{eqnarray} 
for density perturbations and the RGWs respectively. In the above expression $k_0$ is an arbitrarily chosen pivot wavenumber, $n_s$ is the primordial power spectral index for density perturbations, and $n_t$ is the primordial power spectral index for RGWs. $A_s(k_0)$ and $A_t(k_0)$ are normalization coefficients determining the absolute value of the primordial power spectra at the pivot wavenumber $k_0$. The choices of $n_s=1$ and $n_t=0$ correspond to the scale invariant power spectra for density perturbations and gravitational waves respectively.  
The quantity $P_{\mathcal{R}}(k)$ is the primordial power spectrum of the curvature perturbation $\mathcal{R}$ in the comoving gauge, i.e. $P_{\mathcal{R}}(k) = k^3\langle |\mathcal{R}_k|^2\rangle / 2\pi^2$ (see \cite{CAMBnotes} for a detailed exposition). The quantity $P_h(k)$ is the primordial power spectrum of RGWs and gives the mean-square value of the gravitational field perturbations, in a logarithmic interval of the wave-number $k$, at some initial epoch when the wavelenghts of interest are well outside the horizon. 

The relative contribution of density perturbations and gravitational waves is described by the so-called tensor-to-scalar ratio $r$ defined as follows
\begin{eqnarray} \label{r-define}
r(k_0)\equiv \frac{A_t(k_0)}{A_s(k_0)}.
\end{eqnarray}
Note that, in defining the tensor-to-scalar ratio $r$, we have not used any inflationary formulae which relate $r$ with the physical conditions during inflation and the slow-roll parameters (see for example \cite{stein}). Thus, our definition depends only on the power spectral amplitudes of density perturbations and RGWs, and does not assume a particular generating mechanism for these cosmological perturbations.

Assuming that the amplitude of density perturbations $A_s(k_0)$ is known, taking into account the definitions (\ref{gw-spectrum}) and (\ref{r-define}), the power spectrum of the RGW field may be completely characterized by tensor-to-scalar ratio $r$ and the spectral index $n_t$. The RGW amplitude $A_t\left(k_0\right) = r(k_0)A_s(k_0)$ provides us with direct information on the Hubble parameter of the very early universe \cite{GrishchukSolokhin}. More specifically, this amplitude is directly related to the value of the Hubble parameter $H$ at a time when wavelengths corresponding to the wavenumber $k_0$ crossed the horizon \cite{a1,a2a,a2,GrishchukSolokhin,wmap_notation}
\begin{eqnarray}
\nonumber
A_t^{1/2} (k_0)= \left.\frac{\sqrt{2}}{M_{\rm pl}}\frac{H}{\pi}\right|_{k_0/a = H},
\end{eqnarray}
where $M_{\rm pl}=1/\sqrt{8\pi G}$ is the reduced Planck mass.

It is important to point out that, for spectral indices different from the invariant case (i.e., when $n_s\neq1$ and $n_t\neq0$), the definition of the tensor-to-scalar ratio depends on the pivot wavenumber $k_0$.  If we adopt different pivot wavenumber $k_{1}$, the tensor-to-scalar ratio at this new pivot wavenumber $r(k_1)$ is related to original ratio $r(k_0)$ through the following relation (which follows from the definitions (\ref{dp-spectrum}), (\ref{gw-spectrum}) and (\ref{r-define}))
\begin{eqnarray}
\label{r-relation}
r(k_1)=r(k_0) \left(\frac{k_1}{k_0}\right)^{n_t-n_s+1}.
\end{eqnarray}

Let us now turn our attention to CMB. Density perturbations and gravitational waves produce temperature and polarization anisotropies in the CMB characterized by the four angular power spectra $C_{\ell}^{TT}$, $C_{\ell}^{EE}$, $C_{\ell}^{BB}$ and $C_{\ell}^{TE}$ as functions of the multipole number $\ell$. Here $C_{\ell}^{TT}$ is the power spectrum of the temperature anisotropies, $C_{\ell}^{EE}$ and $C_{\ell}^{BB}$ are the power spectra of the so-called $E$ and $B$ modes of polarization (note that, density perturbation do not generate $B$-mode of polarization \cite{a8}), and $C_{\ell}^{TE}$  is the power spectrum of the temperature-polarization cross correlation. In what follows, we shall use the short hand notations $C_{\ell}^{T}$, $C_{\ell}^{E}$, $C_{\ell}^{B} $ and $C_{\ell}^{C}$ to denote these spectra.

In general, the various power spectra $C_{\ell}^{Y}$ (where $Y=T,E,B$ or $C$) can be presented in the following form
\begin{eqnarray}\label{c-sum}
C_{\ell}^{Y}=C_{\ell,s}^{Y}+C_{\ell,t}^{Y},
\end{eqnarray}
where $C_{\ell,s}^{Y}$ is the power spectrum due to the density perturbations (scalar perturbations), and $C_{\ell,t}^{Y}$ is the power spectrum due to RGWs (tensor perturbations).

In the case of RGWs, the various CMB power spectra can be presented in the following form \cite{a10, a11, a12}
\begin{eqnarray}
\label{exact-clxx'}
\begin{array}{c}
C_{\ell,t}^{Y}=(4\pi)^2 \int \frac{dk}{k} P_{h}(k) \left[\Delta^{(T)}_{Y\ell}(k)\right]^2, ~~{\rm for}~Y=T,E,B, \\
C_{\ell,t}^{C}=(4\pi)^2 \int \frac{dk}{k} P_{h}(k) \left[\Delta^{(T)}_{T\ell}(k)\Delta^{(T)}_{E\ell}(k)\right]. 
\end{array}
\end{eqnarray}
Similar expressions hold in the case CMB anisotropies due to density perturbations with a single exception. Density perturbations do not produce the $B$-mode of polarization \cite{a8}. Thus, the CMB power spectra have the form \cite{a10}
\begin{eqnarray}
\label{exact-clxx'DP}
\begin{array}{c}
C_{\ell,s}^{Y}=(4\pi)^2 \int \frac{dk}{k} P_{\mathcal{R}}(k) \left[\Delta^{(S)}_{Y\ell}(k)\right]^2, ~~{\rm for}~Y=T,E, \\
C_{\ell,s}^{C}=(4\pi)^2 \int \frac{dk}{k} P_{\mathcal{R}}(k) \left[\Delta^{(S)}_{T\ell}(k)\Delta^{(S)}_{E\ell}(k)\right]. 
\end{array}
\end{eqnarray}
The transfer functions $\Delta_{Y \ell}^{(S,T)}(k)$ (see \cite{a10, a11, a12} for details) in the above expressions translate the power in the metric fluctuations (density perturbations or gravitational waves) into corresponding CMB power spectrum at an angular scale characterized by multipole $\ell$. In general, these transfer functions are peaked at values $\ell \simeq (1.35 \cdot 10^4 {\rm ~Mpc})\times k$, which is a reflection of the fact that metric fluctuations at a particular linear scale $k^{-1}$ lead to CMB anisotropies predominantly at angular scales $\theta\sim kD$ (where $D$ is the distance to the surface of last scattering). In this work, for numerical evaluation of the various CMB power spectra due to density perturbations and gravitational waves, we use the publicly available CAMB code \cite{camb}.

Since we are primarily interested in the parameters of the RGW field, in the analytical and numerical analysis below we shall work with a fixed cosmological background model. More specifically, we shall work in the framework of $\Lambda$CDM model, and keep the background cosmological parameters fixed at the values determined by a typical model \cite{WMAP3}
\begin{eqnarray}
\label{background}
h=0.732,~\Omega_b
h^2=0.02229,~\Omega_{m}h^2=0.1277,~\Omega_{k}=0,~\tau_{reion}=0.089.
\end{eqnarray} 
Furthermore, for density perturbations, we shall use a model with primordial scalar perturbation power spectrum characterized by an amplitude and spectral index
\begin{eqnarray}
\label{dp} 
A_s=2.3\times
10^{-9},~~n_s=1.0.
\end{eqnarray} 

In light of the above, CMB power spectra produced by RGWs depend on the tensor-to-scalar ratio $r$ and the spectral index $n_t$. In general, this dependence is complicated and requires numerical calculations. For analytical calculations in Section \ref{s3}, we shall use a simple analytical approximation for this dependence (see for example \cite{kosowsky})
\begin{eqnarray}
C_{\ell,t}^{Y} &\simeq& \hat{C}_{\ell,t}^{Y}
\left(\frac{r}{\hat{r}}\right) \left(\frac{\ell}{\ell_0}\right)^{n_t-\hat{n}_t}
\nonumber \\
&=& \hat{C}_{\ell,t}^{Y}\left(\frac{r}{\hat{r}}\right)
\exp\left[(n_t-\hat{n}_t)\ln \left({\ell}/{\ell_0}\right)\right].
\label{appro-clxx'-1}
\end{eqnarray}
Here $\hat{C}_{\ell,t}^Y = C_{\ell,t}^Y(r=\hat{r},n_t=\hat{n}_{t})$ are the spectra calculated for values of tensor-to-scalar ratio and the spectral index fixed at fiducial values $\hat{r}$ and $\hat{n}_{t}$, and $\ell_0$ is the pivot multipole. The approximation (\ref{appro-clxx'-1}) can be further simplified, for values of spectral index $n_t$ sufficiently close to the fiducial value $\hat{n}_{t}$ (such that $(n_t-\hat{n}_{t})\ln\left(\ell/\ell_0\right)\ll 1$)
\begin{eqnarray}
\label{appro-clxx'-2}
C_{\ell,t}^{Y} &\simeq& \hat{C}_{\ell,t}^{Y}
\left(\frac{r}{\hat{r}}\right) \left[1+ (n_t-\hat{n}_{t})\ln\left({\ell}/{\ell_0}\right)\right].
\end{eqnarray}

\begin{figure}[t]
\centerline{\includegraphics[width=18cm,height=12cm]{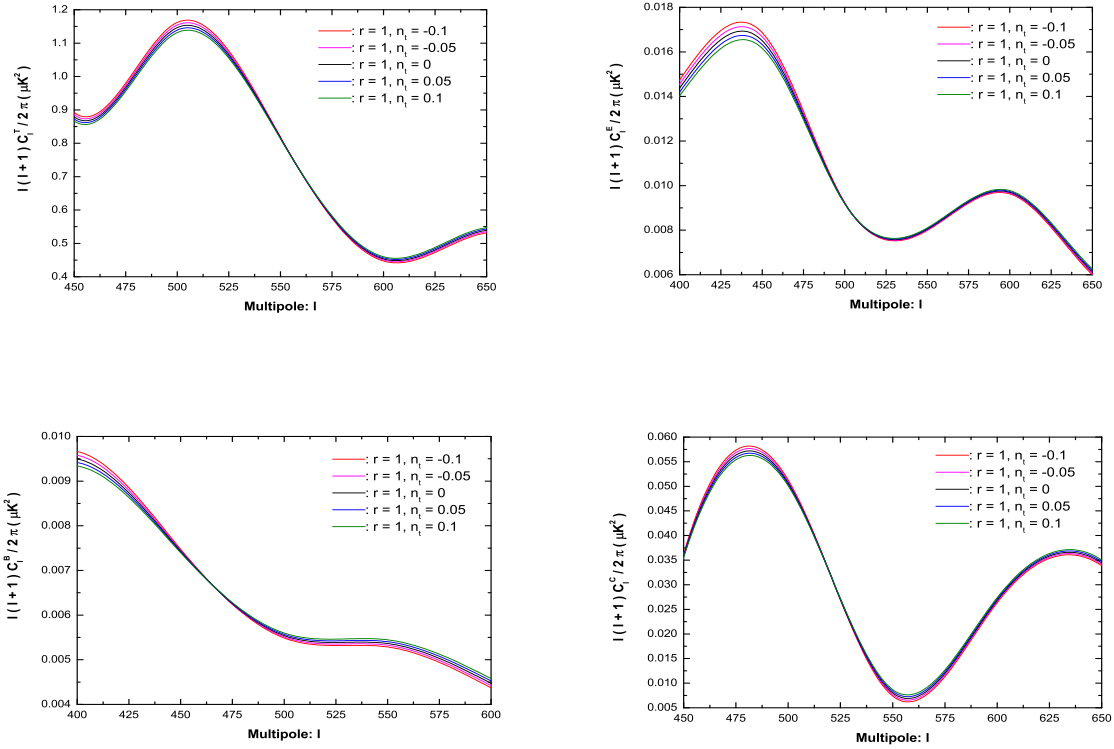}}
\caption{The CMB power spectra due to RGWs for various values of the spectral index $n_t$: $C_{\ell,t}^{T}$ (left upper panel),  $C_{\ell,t}^{E}$ (right upper panel),  $C_{\ell,t}^{B}$ (left lower panel),  $C_{\ell,t}^{C}$ (right upper panel). The pivot wavenumber is chosen $k_0=0.05$Mpc$^{-1}$ (in all the panels), and the power spectra are shown for multipoles around the value of the corresponding pivot multipole $\ell\sim 500$.}\label{figurea1}
\end{figure}

The pivot multipole $\ell_0$ is closely related to the pivot wavenumber $k_0$.  The approximation (\ref{appro-clxx'-1}) can be derived from (\ref{gw-spectrum}) and (\ref{exact-clxx'}) under the assumption that the wavenumber $k$ and multipole $\ell$ are linearly related, i.e.~$k/k_0 \sim \ell/\ell_0$. This assumption is justified due to the peaked nature of transfer functions $\Delta_{Y,\ell}^{(T)}(k)$ entering (\ref{exact-clxx'}). Numerical evaluations show that the pivot multipole is related to pivot wavenumber by 
\begin{eqnarray}
\label{k-l}
\ell_0 \approx k_0\times 10^{4}{\rm Mpc}.
\end{eqnarray}
For illustration, in FIG. \ref{figurea1} we plot the power spectra $C_{\ell,t}^{Y}$ for different value of the spectral index $n_t$. The pivot wavenumber is taken to be $k_0=0.05$Mpc$^{-1}$. As expected, in all the panels the spectra with different values of $n_t$ converge at $\ell\simeq 500$, which is consistent with the prediction of the relation (\ref{k-l}).

The CMB power spectra $C_{\ell}^{Y}$ are theoretical constructions determined by ensemble averages over all possible realizations of the underlying random process. However, in real CMB observations, we only have access to a single sky, and hence to a single realization. In order to obtain information on the power spectra from a single realization, it is required to construct estimators of power spectra. In order to differentiate the estimators from the actual power spectra, we shall use the notation $D_{\ell}^Y$ to denote the estimators while retaining the notation $C_{\ell}^Y$ to denote the power spectrum. It is important to keep in mind that the estimators $D_{\ell}^Y$ are constructed from observational data, while the power spectra $C_{\ell}^Y$ are theoretically predicted quantities. The probability density functions (pdfs) for the estimators are described in detail in Appendix \ref{appendix-a}. In what follows, we shall require the data from all the power spectral estimators, i.e.~$D_{\ell}^Y$ for $Y=T,E,B {\rm ~and~}C$. Let us denote this set of estimators (which we shall sometimes refer to as the sample) as
\begin{eqnarray}
\{D_{\ell}^{Y}\} \equiv \{D_{\ell}^{Y}|Y=C,T,E,B;\ell=2,3,\cdot\cdot\cdot,\ell_{\max}\}.
\nonumber
\end{eqnarray}
To simulate an experiment, we shall randomly draw a data set $\left\{D_{\ell}^Y\right\}$ from the pdf (\ref{joint}). In calculating the pdf (\ref{joint}), along with parameters given in (\ref{background}) and (\ref{dp}),
we set the value of the RGW parameters as
\begin{eqnarray}
\label{input}
r=\hat{r},~~~n_t=\hat{n}_t.
\end{eqnarray}
We shall refer to $\hat{r}$ and $\hat{n}_t$ as the parameters of the input model. 

For analytical evaluations in Section \ref{s3}, we shall work with Gaussian approximation to the exact pdfs (\ref{joint}) for the estimators $\left\{D_{\ell}^Y\right\}$. The Gaussian approximation is characterized by corresponding mean values and standard deviations \cite{ours}
\begin{eqnarray}
\label{mean-variance}
\begin{array}{c}
\langle D_{\ell}^{Y}\rangle 
=C_{\ell}^{Y},~~(Y=T,E,B,C), \\
\sigma_{D_{\ell}^{Y}}
=\sqrt{\frac{2}{(2\ell+1)f_{\rm sky}}}(C_{\ell}^{Y}+N_{\ell}^{Y}W_{\ell}^{-2}),
~~(Y=T,E,B), \\
\sigma_{D_{\ell}^{C}}=
\sqrt{\frac{(C_{\ell}^C)^2
+(C_{\ell}^T+N_{\ell}^TW_{\ell}^{-2})(C_{\ell}^E
+N_{\ell}^EW_{\ell}^{-2})}{(2\ell+1)f_{\rm sky}}}.
\end{array}
\end{eqnarray}
Note that, the above expressions for mean values and standard deviations follow from the exact pdfs considered in Appendix \ref{appendix-a}. In the above expression, $N_\ell^Y$ are the noise power spectra, $f_{\rm sky}$ is the cut sky factor, and $W_\ell$ is the window function.

In the case of the Planck mission \cite{Planck}, considering the channel at 143GHz (which has the lower foreground level and lowest noise power spectra) the noise power spectra, the cut sky factor and the window function are given by \cite{Planck} (see \cite{ours,ours2} for further explanations)
\begin{eqnarray}
\label{planck}
\begin{array}{c}
 N_{\ell}^{T}=1.53\times 10^{-4} {\rm \mu K}^2,~~N_{\ell}^{E}=N_{\ell}^{B}=5.58\times 10^{-4} {\rm \mu
 K}^2,\\ f_{\rm sky}=0.65,~~ W_{\ell}=\exp\left[-\frac{\ell(\ell+1)}{2}\frac{\theta^2_{\rm FWHM}}{8\ln 2}\right],
\end{array}
\end{eqnarray}
 where $\theta_{\rm FWHM}=7.1'$ is the full width at half maximum of the Gaussian beam. 
 
 In this paper, along with predictions for Planck, we shall consider an idealized situation with no instrumental noise, full sky coverage and an idealized window function $W_\ell=1$. For this case, we shall assume that the only source of noise comes from contribution of cosmic lensing to the $B$-mode of polarization. In this case the noise spectrum for the $B$-mode is close to white with a value $N_{\ell}^B\simeq 2\times 10^{-6}\mu$K$^2$ \cite{wuran4,lensing1}. A number of works have discussed methods to subtract the lensing $B$-mode signal (see for example \cite{lensing1,lensing2}). In \cite{lensing2}, the authors claimed that a reduction in lensing power by a factor of 40 is possible using approximate iterative maximum-likelihood method. For this reason, as a further idealized but feasible scenario, we shall also consider the case with reduced cosmic lensing noise $N_{\ell}^B\simeq 5\times 10^{-8}\mu$K$^2$. Thus in the two described examples the noises are 
 \begin{eqnarray}
 \label{ideal1}
 \begin{array}{c}
N_{\ell}^{B}({\rm lensing})=2\times 10^{-6} {\rm \mu
 K}^2,~~N_{\ell}^{B}({\rm reduced~lensing})=5\times 10^{-8} {\rm \mu
 K}^2;\\
  N_{\ell}^{T}=N_{\ell}^{E}=0;~
 ~f_{\rm sky}=1;~W_{\ell}=1.
 \end{array}
 \end{eqnarray} 
Note that, in addition to the instrumental noises and lensing noise, various foregrounds, such as the the synchrotron and dust, significantly contaminate the CMB signal. However, it is hoped that, using multifrequency observations together with ingenious foreground subtraction techniques, future experiments would be able approach the ideal limit of expression (16) (see for instant \cite{foreground}).

Before proceeding, let us briefly mention the notational conventions used in this paper. The $star$ superscript denotes the quantities evaluated at the best pivot multipole $\ell^{*}_t$. The $hat$ superscript indicates the parameters of the fixed (input) cosmological model, that are used to generate the simulated observational data.
The summation (product) symbols with subscript $\ell$ or $Y$ indicate summation (product) for $\ell=2,...,\ell_{max}$ and $Y=C,T,E,B$ respectively. In numerical evaluation we set $\ell_{max}=1000$.


\section{analytical approximation\label{s3}}

In this section, we shall derive analytical expressions for the estimation of the parameters $r$ and $n_t$, the associated uncertainties $\Delta r$ and $\Delta n_t$ and the correlation between these parameters. We will show the existence and explain the significance of the best pivot multipole $\ell_t^*$. Introducing the tensor-to-scalar ratio $r^*$ defined at the best pivot multipole, we shall show that this parameter can be determined with the smallest possible uncertainty, and is not correlated with the spectral index $n_t$. Based on this analysis, we shall discuss the signal-to-noise ratio and detection possibilities for various CMB experiments.


\subsection{Approximation for the likelihood function\label{s3.1}}

In order to estimate the parameters $r$ and $n_t$ characterizing the RGW background, we shall use an analysis based on the likelihood function \cite{bayesian,cosmomc}. The likelihood function is just the probability density function of the observational data considered as a function of the unknown parameters (which are $r$ and $n_t$ in our case). Up to a constant, independent of its arguments, the likelihood function is given by 
\begin{eqnarray}
\nonumber
\mathcal{L} = \prod_{\ell} f(D_{\ell}^{C},D_{\ell}^{T},D_{\ell}^{E},D_{\ell}^{B}),
\end{eqnarray}
where the function $f(D_{\ell}^{C},D_{\ell}^{T},D_{\ell}^{E},D_{\ell}^{B})$ is explained in detail in Appendix \ref{appendix-a}. 

For analysis in this section, we shall use a Gaussian function to approximate the pdf of the individual estimator $D_{\ell}^{Y}$, and ignore any possible correlation between different estimators. In this case the approximate likelihood function can be written as (see \cite{ours2} for details)
\begin{eqnarray}
\label{a1a}
 -2\ln \mathcal{L} =
 \sum_{\ell}\sum_{Y}
 \left(\frac{D_{\ell}^{Y}-C_{\ell}^{Y}}{{\sigma}_
{ D_{\ell}^{Y}}}\right)^2.
\end{eqnarray}
The parameters $r$ and $n_t$ enter the above expression through the quantities $C_{\ell}^{Y}$ and $\sigma_{D_{\ell}^Y}$. In our analytical considerations we shall make a further simplification. We shall assume that $\sigma_{D_{\ell}^Y}$ entering (\ref{a1a}) is weekly dependent on parameters $r$ and $n_t$ and assume $\sigma_{D_{\ell}^Y} = \hat{\sigma}_{D_{\ell}^Y}$ (for a justification of this assumption see \cite{ours,ours2}). With this assumption, the likelihood function can be rewritten as follows
\begin{eqnarray}
\label{a1}
 -2\ln \mathcal{L} =
 \sum_{\ell}\sum_{Y}
 \left(\frac{D_{\ell}^{Y}-C_{\ell}^{Y}}{\hat{\sigma}_
{ D_{\ell}^{Y}}}\right)^2.
\end{eqnarray}

In the likelihood analysis, we shall assume that the value of the sought parameter $n_t$ is sufficiently close to the input value $\hat{n}_t$. In this case, inserting (\ref{appro-clxx'-2}) into (\ref{a1}), using (\ref{c-sum}), we can rewrite the likelihood in the form
\begin{eqnarray}\label{a5}
-2\ln \mathcal{L}=\sum_{\ell}\sum_{Y}\left\{ a_{\ell}^Y
\left[\left(\frac{r}{\hat{r}}\right)\left(1+(n_t-\hat{n}_{t})b_{\ell}\right)\right]-d_{\ell}^{Y}\right\}^2,
\end{eqnarray}
where
\begin{eqnarray}\label{a6}
a_{\ell}^Y\equiv \frac{\hat{C}_{\ell,t}^{Y}}{\hat{\sigma}_{
D_{\ell}^Y}},~~b_{\ell}\equiv\ln\left(\frac{\ell}{\ell_{0}}\right),~~
d_{\ell}^Y\equiv \frac{D_{\ell}^{Y}-C_{\ell,s}^Y}{
\hat{\sigma}_{D_{\ell}^Y}}.
\end{eqnarray}
Note that, in the above expression, the dependence on the data (on the estimators $D_{\ell}^Y$) is solely contained in the term $d_{\ell}^Y$. Furthermore, $a_{\ell}^Y$, $b_{\ell}$ and $d_{\ell}^{Y}$ are independent of the RGW parameters $r$ and $n_t$. The dependence on $r$ and $n_t$ takes a particularly simple form and is contained within the square brackets on the right side in (\ref{a5}).

In order to proceed, it is convenient to introduce new variables 
\begin{eqnarray}
\label{a7}
\xi\equiv r/\hat{r} ~~\zeta\equiv(n_t-\hat{n}_{t})(r/\hat{r}),
\end{eqnarray}
in place of $r$ and $n_t$. In terms of these variables, the likelihood (\ref{a5}) can be simplified as
 \begin{eqnarray}\label{k1}
 -2\ln \mathcal{L}=\sum_{\ell}\sum_{Y}\left[a_{\ell}^Y
 (\xi+\zeta b_{\ell})-d_{\ell}^{Y}\right]^2.
 \end{eqnarray}
Note that, the dependence on the sought for parameters $r$ and $n_t$, in the above expression, is contained in the variables $\xi$ and $\zeta$. After a straight forward manipulations (\ref{k1}) can be rewritten as
 \begin{eqnarray}
 \nonumber
 \begin{array}{c}
 -2\ln \mathcal{L}=\xi^2(\sum_{\ell}
 \sum_{Y}a_{\ell}^{Y2})+
 \zeta^2(\sum_{\ell}\sum_{Y}(a_{\ell}^{Y}b_{\ell})^2)
 +2\xi\zeta(\sum_{\ell}\sum_{Y}a_{\ell}^{Y2}b_{\ell}) \\
 -2\xi(\sum_{\ell}\sum_{Y}a_{\ell}^{Y}d_{\ell}^{Y})
 -2\zeta(\sum_{\ell}\sum_{Y}a_{\ell}^{Y}d_{\ell}^{Y}b_{\ell})
 +\sum_{\ell}\sum_{Y} d_{\ell}^{Y2}.
 \end{array}
 \end{eqnarray}
This expression can be rewritten as of 
\begin{eqnarray}
\label{a8}
\begin{array}{c}
 -2\ln \mathcal{L}=(\sum_{\ell}\sum_{Y} a_{\ell}^{Y2}) 
 \left(\xi-\frac{\sum_{\ell}\sum_{Y}a_{\ell}^Y
 d_{\ell}^Y}{\sum_{\ell}\sum_{Y} a_{\ell}^{Y2}}\right)^2 
 + (\sum_{\ell}\sum_{Y}(a_{\ell}^{Y}b_{\ell})^2)
 \left(\zeta-\frac{\sum_{\ell}\sum_{Y}
 a_{\ell}^{Y}b_{\ell}d^Y_{\ell}}{\sum_{\ell}
 \sum_{Y}(a_{\ell}^{Y}b_{\ell})^2}\right)^2
\\
+2\xi\zeta(\sum_{\ell}\sum_{Y}a_{\ell}^{Y2}b_{\ell})+C,
\end{array}
\end{eqnarray}
where $C$ is a constant, independent of $r$ and $n_t$. This constant is responsible for the overall normalization of the likelihood function and will not participate in estimation of parameters. In the following subsection we shall use the approximation (\ref{a8}) for estimating the parameters $r$ and $n_t$.


\subsection{Posterior pdf and the best pivot multipole $\ell_t^*$\label{s3.2}}

\subsubsection{Posterior pdf\label{s3.2.2}}

The constraint on the parameters $r$ and $n_t$, are determined by the posterior probability density function $P(r,n_t)$. This posterior pdf is related to the likelihood function $\mathcal{L}$ by \cite{bayesian,cosmomc}
\begin{eqnarray}
\label{k3}
P(r,n_t)=f(r,n_t)\mathcal{L},
\end{eqnarray}
where $f(r,n_t)$ is the prior probability density function of the parameters $r$ and $n_t$. In this paper, we adopt a flat prior, i.e.
\begin{eqnarray}
\label{k4}
f(r,n_t)=1.
\end{eqnarray}
Thus, in this case, the posterior pdf $P(r,n_t)$ becomes equal to the likelihood. Using the approximation (\ref{a8}) for the likelihood, we obtain 
\begin{eqnarray}
-2\ln P(r,n_t)&=&(\sum_{\ell}\sum_{Y} a_{\ell}^{Y2}) \left(\xi-\frac{\sum_{\ell}\sum_{Y}a_{\ell}^Y
d_{\ell}^Y}{\sum_{\ell}\sum_{Y} a_{\ell}^{Y2}}\right)^2 + (\sum_{\ell}\sum_{Y}(a_{\ell}^{Y}b_{\ell})^2)
\left(\zeta-\frac{\sum_{\ell}\sum_{Y}a_{\ell}^{Y}b_{\ell}d^Y_{\ell}}{\sum_{\ell}\sum_{Y}(a_{\ell}^{Y}b_{\ell})^2}\right)^2
\nonumber \\
&& +2\xi\zeta(\sum_{\ell}\sum_{Y}a_{\ell}^{Y2}b_{\ell})+C.
\label{k6}
\end{eqnarray}
The parameters $r$ and $n_t$ enter the above expression through the variables $\xi$ and $\zeta$. For this reason it is convenient to firstly consider the posterior pdf for variables $\xi$ and $\zeta$. It will be seen that the posterior pdf for these variables will have a particularly simple form, namely a bivariate normal function. The posterior pdf $P(\xi,\zeta)$ is related to $P(r,n_t)$ in the following manner
\begin{eqnarray}
\label{k7}
P(\xi,\zeta)=\left|\frac{\partial(r,n_t)}{\partial(\xi,\zeta)}
\right|P(r,n_t)=\frac{\hat{r}}{\xi}P(r,n_t),
\end{eqnarray}
where $\left|\frac{\partial(r,n_t)}{\partial(\xi,\zeta)}\right|$ denotes the Jacobian of the transformation between the two sets of variables calculable from (\ref{a7}). For simplicity and clarity, let us firstly consider the constraints on the parameters $\xi$ and $\zeta$. Following this, we shall return to the discussion on $r$ and $n_t$ using relation (\ref{k7}).

Introducing the notations
\begin{eqnarray}
\label{kk9}
\begin{array}{c}
\xi_p\equiv {\left(\sum_{\ell}\sum_{Y}a_{\ell}^Y
d_{\ell}^Y\right)}/{\left(\sum_{\ell}\sum_{Y} a_{\ell}^{Y2}\right)},~
\xi_s\equiv 1/\sqrt{\sum_{\ell}\sum_{Y} a_{\ell}^{Y2}},
\\
\zeta_p\equiv {\left(\sum_{\ell}\sum_{Y}a_{\ell}^Y
d_{\ell}^Y b_{\ell}\right)}/{\left(\sum_{\ell}\sum_{Y}(a_{\ell}^Y b_{\ell})^2\right)},~
\zeta_s\equiv 1/\sqrt{\sum_{\ell}\sum_{Y}
(a_{\ell}^{Y}b_{\ell})^2}.
\end{array}
\end{eqnarray}
From (\ref{k6}) and (\ref{k7}), we obtain that expression for the posterior pdf of $\xi$ and $\zeta$ in the form
 \begin{eqnarray}\label{kk8}
 P(\xi,\zeta)=\frac{\hat{r}e^C}{\xi}\exp\left[-\frac{(\xi-\xi_p)^2}{2(\xi_s)^2}\right]
 \exp\left[-\frac{(\zeta-\zeta_p)^2}{2(\zeta_s)^2}\right]\exp\left[-\xi\zeta(\sum_{\ell}\sum_{Y}a_{\ell}^{Y2}b_{\ell})\right].
 \end{eqnarray}

\subsubsection{Best pivot multipole $\ell_t^*$\label{s3.2.1}}

Let us concentrate on the posterior pdf (\ref{kk8}). As can be seen, there is a non-vanishing correlation between the parameters $\xi$ and $\zeta$ in the case when $\sum_{\ell}\sum_{Y}a_{\ell}^{Y2}b_{\ell}\neq 0$. From the definition (\ref{a6}), it follows that the terms $b_{\ell}$ depend on the arbitrarily chosen pivot multipole $\ell_0$ (corresponding to the pivot wavenumber $k_0$ through the relation (\ref{k-l})). For this reason, we can select the pivot multipole $\ell_0=\ell^*_t$ so as to require
\begin{eqnarray}
\label{a9}
\sum_{\ell}\sum_{Y}a_{\ell}^{Y2}b_{\ell}^*=0,
\end{eqnarray}
where $\left. b_{\ell}^*\equiv b_{\ell}\right|_{\ell_0=\ell_t^*}$.
With this choice of pivot multipole $\ell_t^*$, and the corresponding pivot wavenumber $k_t^*$, the variables $\xi(k_t^*)$ and $\zeta(k_t^*)$ will have no correlation. We shall refer to this pivot multipole $\ell_t^*$ as the Best Pivot Multipole number. From definitions (\ref{a6}) of $a_{\ell}^{Y}$ and $b_{\ell}$, along with expression (\ref{mean-variance}), it follows that the precise numerical value of the best pivot multipole number $\ell^*_t$ depends on the input cosmological model characterized by the (\ref{background}), (\ref{dp}) and (\ref{input}), as well as the specifics of the CMB experiment characterized by noise power spectra, cut sky factor and window function. We shall discuss this dependence in more detail below.

Setting the value of the pivot multipole $\ell_0=\ell_t^*$, so as to satisfy (\ref{a9}), we arrive at a simplified form for the posterior pdf
\begin{eqnarray}
\label{k8}
P(\xi^*,\zeta^*)=(\hat{r}e^C)\frac{1}{\xi^*}\exp\left[-\frac{(\xi^*-\xi_p^*)^2}{2(\xi_s^*)^2}\right]
\exp\left[-\frac{(\zeta^*-\zeta_p^*)^2}{2(\zeta_s^*)^2}\right].
\end{eqnarray}
As a reminder let us point out that, in the above expression, as well as in what follows, we have used notations $ r^*$, $\xi^*$, $\zeta^*$ and $b^*_{\ell}$ to denote the corresponding quantities calculated for the pivot multipole chosen at the best pivot multipole value $\ell^*_t$. Note that for the spectral index of RGWs we shall retain the notation $n_t$, since it does not depend on the choice of the pivot multipole.

\subsubsection{Constraints on parameters $\xi^*$ and $\zeta^*$\label{s3.2.3}}

Equipped with the posterior pdf (\ref{k8}), let us analyze the uncertainties in determining the parameters $\xi^*$ and $\zeta^*$. For simplicity of analysis we shall assume 
\begin{eqnarray}
\label{problem}
\xi_p^*\gg
\xi_s^*.
\end{eqnarray}
In Section \ref{s3.3.2}, we shall show that this constraint corresponds to a condition that the signal-to-noise ratio is large, i.e.~$S/N\gg1$. Taking into account this condition,  the posterior function (\ref{k8}) may be further approximated in the following manner
\begin{eqnarray}\label{kk10}
P(\xi^*,\zeta^*)\simeq\frac{\hat{r}e^C}{\xi_p^*}\exp\left[-\frac{(\xi^*-\xi_p^*)^2}{2(\xi_s^*)^2}\right]
\exp\left[-\frac{(\zeta^*-\zeta_p^*)^2}{2(
\zeta_s^*)^2}\right].
\end{eqnarray}
Note that, in the above expression, the factor in front of the exponent $\hat{r}e^C/\xi_p^*$ now becomes a constant independent of $\xi^*$ and $\zeta^*$. Thus, the posterior pdf $P(\xi^*,\zeta^*)$ becomes a bivariate normal (Gaussian) function for variables $\xi^*$ and $\zeta^*$. The position of the maximum and the standard deviation associated with the posterior pdf $P(\xi^*,\zeta^*)$ are given by
\begin{eqnarray}
\label{a10}
\xi^*_{ML}=\xi^*_{p},~~
\Delta{{\xi^*}}=\xi_s^*,~~ \zeta^*_{ML}=\zeta^*_{p},~~
\Delta{{\zeta^*}}=\zeta_s^*.
\end{eqnarray} 
In the above expression, subscript ``$ML$" stands for ``maximum-likelihood", since the maximum of the posterior pdf coincides with that of the likelihood function due to (\ref{k3}) and (\ref{k4}). Following the maximum likelihood parameters estimation procedure, we shall accept the values $\xi^*_{ML}$ and $\zeta^*_{ML}$ as the estimators for the corresponding quantities $\xi^*$ and $\zeta^*$. It is worth mentioning that, for the posterior pdfs considered in this work, the maximum-likelihood values coincide with the mean values of the corresponding posterior pdfs. It is worth mentioning that, the assumption $\xi_p^*\gg\xi_s^*$ (which was used to derive the pdf (\ref{kk10})) is equivalent to the requirement $\xi^*_{ML}\gg \Delta{{\xi^*}}$.

Proceeding further, we can calculate the correlation coefficient for variables $\xi^*$ and $\zeta^*$. Let us firstly define the covariance in the following manner
\begin{eqnarray}
{\rm cov}(x,y) \equiv \overline{\left(x-\overline{x}\right)\left(y-\overline{y}\right)},
\label{covariancedef}
\end{eqnarray}
where the overline indicates averaging over the corresponding posterior pdf. The correlation coefficient can now be explicitly calculated to give
\begin{eqnarray}
\label{rho-define}
\rho_{(\xi^*,\zeta^*)}\equiv
\frac{{\rm cov}(\xi^*,\zeta^*)}{\sqrt{{\rm cov}(\xi^*,\xi^*){\rm cov}(\zeta^*,\zeta^*)}} = 0,
\end{eqnarray}
as expected, the correlation between the variables $\xi^*$ and $\zeta^*$ vanishes.

Taking into account (\ref{kk9}), we find that the mean values $\overline{\xi^*}$ (also $\xi^*_{ML}$) and $\overline{\zeta^*}$ (also $\zeta^*_{ML}$) depend on data $\{D_{\ell}^{Y}\}$ through quantities $d_{\ell}^Y$. However, in our approximation, the standard deviations  $\Delta{{\xi^*}}$ and $\Delta{{\zeta^*}}$ are independent of the data. They depend on the input cosmological model determined by (\ref{background}), (\ref{dp}), (\ref{input}), along with noises, cut sky factor and the window function characterizing the CMB experiment.


\subsection{Constraints on parameters $r^*$ and $n_t$\label{s3.3} \label{s3.3.1}}

Let us now return to the parameters of our direct interest, namely the tensor-to-scalar ratio (determined at the best pivot wavenumber) $r^*$ and RGW primordial spectral index $n_t$. These parameters are related to $\xi^*$ and $\zeta^*$ through relations
\begin{eqnarray}
\label{cd4}
r^*=\hat{r}^*\xi^*,~~n_t=\hat{n}_{t}+\zeta^*/\xi^*,
\end{eqnarray}
which follow from (\ref{a7}). Taking into account the fact that the quantity $\xi^*$ is peaked at $\xi_{ML}^*$ which is sufficiently close to the input value $\hat{\xi}^*$, and $\Delta\xi^*/\xi_{ML}^*\ll 1$, we can approximate the quantity $\xi^*$ in the expression for $n_t$ (see the second formula in (\ref{cd4})) with $1$. Thus (\ref{cd4}) can be written as
\begin{eqnarray}
\label{c4}
r^*=\hat{r}^*\xi^*,~~n_t\simeq\hat{n}_{t}+\zeta^*.
\end{eqnarray}

Using (\ref{k7}), the posterior pdf for $r^*$ and $n_t$ is related to $P(\xi^*,\zeta^*)$ in the following way
\begin{eqnarray}
\label{posterior2}
P(r^*,n_t)=\frac{\xi^*}{\hat{r}^*}P(\xi^*,\zeta^*)\simeq \frac{\xi_p^*}{\hat{r}^*}P(\xi^*,\zeta^*).
\end{eqnarray}
Note that, in the current approximation, the pdf for variables $r^*$ and $n_t$ has a bivariate normal form.

Based on the above pdf (\ref{posterior2}), we can now evaluate the maximum likelihood estimators, standard deviations, and the correlation coefficient for the variables $r^*$ and $n_t$. For the maximum likelihood values we get
\begin{eqnarray}
\label{meanvaluernt}
\begin{array}{l}
{r^*_{ML}} = \hat{r}^*{\xi^*_{ML}} = \hat{r}^* \frac{\sum_{\ell}\sum_{Y}a_{\ell}^Y
 d_{\ell}^Y}{\sum_{\ell}\sum_{Y} a_{\ell}^{Y2}}, \\
{n_t}_{ML} \simeq \hat{n}_t + {\zeta^*_{ML}} =  \hat{n}_{t}+\frac{\sum_{\ell}\sum_{Y}a_{\ell}^Y
 d_{\ell}^Y b^*_{\ell}}{\sum_{\ell}\sum_{Y}(a_{\ell}^Y b^*_{\ell})^2}. 
\end{array}
\end{eqnarray}
The standard deviation are given by
\begin{eqnarray}
\label{k11}
\begin{array}{l}
\Delta{r^*} = \hat{r}^*\Delta{\xi^*} = \hat{r}^*/\sqrt{\sum_{\ell}\sum_{Y} a_{\ell}^{Y2}}, \\
\Delta{n_t} \simeq \Delta\zeta^*= 1/\sqrt{\sum_{\ell}\sum_{Y}
 (a_{\ell}^{Y}b^{*}_{\ell})^2}. 
\end{array}
\end{eqnarray}
Finally, it can be shown that the correlation between $r^*$ and $n_t$ vanishes 
\begin{eqnarray}
\label{k14}
\rho_{(r^*, n_t)}\equiv \frac{{\rm cov} (r^*, n_t)}{\sqrt{{\rm cov} (r^*, r^*){\rm cov} (n_t, n_t)}} = 0.
\end{eqnarray}
It is interesting to point out that these results are consistent with the results in \cite{ours2}. The constraints, presented in this paper, on the tensor-to-scalar ratio $r$ are exactly the same as those in \cite{ours2}. Unlike the analysis in the current paper, \cite{ours2} works with a single free parameter $r$ and does not consider $n_t$ as an independent free parameter.


\subsection{Constraints on parameter $r$\label{s3.4} \label{s.3.4.1} \label{s3.4.2}}

In Subsection \ref{s3.3}, using the posterior probability function $P(r^*,n_t)$, we have investigated the tensor-to-scalar ratio $r^*$ and the spectral index $n_t$ defined at the best pivot wavenumber $k_t^*$ (corresponding to the best pivot multipole $\ell_t^*$). We can now proceed to the analysis of the tensor-to-scalar ratio $r$, defined at an arbitrary pivot wavenumber $k_0$, and determine the possible constraints on this parameter.

From the Eq.(\ref{r-relation}), we can express the tensor-to-scalar ratio $r$, in terms of $r^*$, $k_t^*$ and the spectral indices $n_t$ and $n_s$, in the following form
\begin{eqnarray}\label{j3}
\ln r=\ln r^*+n_t\ln(k_0/k_t^*)+(1-n_s)\ln(k_0/k_t^*).
\end{eqnarray}
It can be seen that, for a fixed value of the spectral index $n_s$ (see (\ref{dp})), $r$ depends on the parameters $r^*$ and $n_t$. Thus, the properties of $r$ can be determined using the posterior pdf $P(r^*,n_t)$, which was analyzed in detail in Section \ref{s3.3}. In the case $k_0\neq k_t^*$ it is more illustrative to consider the variable $\ln r$ instead of the variable $r$. For this reason, when dealing with the maximum likelihood estimators of tensor-to-scalar ratio defined at pivot scale different from the best-pivot scale, we shall use the corresponding logarithms 
\begin{eqnarray}
\label{overliner}
{\ln r}_{ML} = \ln{r^*_{ML}}+{n_t}_{ML}\ln(k_0/k_t^*)+(1-n_s)\ln(k_0/k_t^*),
\end{eqnarray}
where ${r^*_{ML}}$ and ${n_t}_{ML}$ are the maximum likelihood estimators expressible in terms of the input parameters $\hat{r}^*$, $\hat{n}_t$ and the data $\{D_{\ell}^X\}$ (see (\ref{meanvaluernt})). The uncertainty of $r$ can be expressed in terms of the uncertainties $\Delta {r^*}$ and $\Delta{n_t}$ determined in (\ref{k11}), leading to the following expression
\begin{eqnarray}
\Delta{\ln r} 
&\simeq &
\sqrt{(\Delta {r^*}/\hat{r}^*)^2+(\ln (k_0/k_t^*)\Delta{n_t})^2}, 
\nonumber \\
&=& 
\sqrt{(\xi_s^*)^2+(\ln(k_0/k_t^*)\zeta_s^*)^2}.
\label{j5}
\end{eqnarray}
The quantities $\xi^*_s$ and $\zeta^*_s$, entering the above expression, can be expressed through CMB power spectra due to RGWs $C_{\ell,t}^X$ using (\ref{a6}) and (\ref{kk9}).

From (\ref{j5}) it follows that $\Delta r/r \gtrsim \Delta r^*/r^*$, with the equality holding for $k_0\rightarrow k_t^*$. Thus, the smallest uncertainty on tensor-to-scalar ratio $r$ is achieved for the choice of the pivot scale $k_0=k_t^*$. This justifies the title ``best" pivot wavenumber for $k_t^*$. For a choice of pivot wavenumber $k_0\neq k_t^*$ the uncertainty in determining $r$ becomes larger due to the uncertainty in determining the spectral index  $n_t$.

Although, as was shown in Subsection {\ref{s3.3}}, the quantities $r^*$ and $n_t$ are uncorrelated, this is not true for the quantities $r$ and $n_t$ in general. In order to describe the correlation between $r$ and $n_t$, it is convenient to introduce the correlation coefficient
\begin{eqnarray}
\label{lnr-rho}
\rho_{(n_t,\ln r)}\equiv\frac{{\rm cov}(n_t, \ln r)}{\sqrt{{\rm cov}(n_t, n_t)
{\rm cov}(\ln r, \ln r)}},
\end{eqnarray}
where the notation ${\rm cov}(\cdot,\cdot)$ for the covariance was defined in (\ref{covariancedef}). Using this definition, along with (\ref{j3}) and (\ref{k14}), the terms entering the above expression can be evaluated as
\begin{eqnarray}
\nonumber 
\begin{array}{l}
{\rm cov}(\ln r, \ln r)=(\Delta {\ln r})^2, \\
{\rm cov}(n_t, n_t)=(\Delta {n_t})^2, \\ 
{\rm cov}(n_t, \ln r)= {\rm cov}(n_t, \ln r^*)+(\ln(k_0/k_t^*)) {\rm cov}(n_t, n_t) = (\ln(k_0/k_t^*)) (\Delta n_t)^2. \\
\end{array}
\end{eqnarray}
Taking into account (\ref{k11}), the correlation coefficient can be presented in the following form
 \begin{eqnarray}\label{ff2}
 \rho_{(n_t,\ln r)}
 = \sqrt{\frac{{\zeta_s^*}^2\left(\ln(k_0/k_t^*)\right)^2}{{{\zeta_s^*}^2\left(\ln(k_0/k_t^*)\right)^2+{{\xi_s^*}}^2}}}.
 \end{eqnarray}
As expected, for choice  the $k_0= k_{t}^*$, i.e.~when the pivot wavenumber is chosen at the value of the best pivot wavenumber, the correlation between $r$ and $n_t$ vanishes. On the other hand, for $|\ln(k_0/k_t^*)|\gg1$, i.e. for values of the pivot wavenumber significantly different from the best pivot wavenumber, the correlation coefficient approaches unity, implying a strong correlation between $r$ and $n_t$.


\subsection{Statistical properties of maximum likelihood estimators \label{section-manyrealizations}}

The exact numerical values of the maximum likelihood ($ML$) estimators $\xi_{ML}^*$, $\zeta^*_{ML}$, $r^*_{ML}$, ${n_t}_{ML}$ and $\ln r_{ML}$ discussed in the previous subsections depend on the CMB data $\{D_{\ell}^{Y}\}$. Since the set $\{D_{\ell}^{Y}\}$ is a single realization of an underlying random process characterized by the pdf (\ref{joint}), the precise values of the maximum likelihood estimators will depend on this realization. For this reason, it is instructive to analyze the distribution of these maximum likelihood estimators in various realizations of the underlying random process specified by the pdf for estimators of the CMB power spectrum $\{D_{\ell}^{X}\}$. Heuristically speaking, the mean value of this distribution characterizes the typical value for the $ML$ estimators that we are likely to observe (for a specific input cosmological model), while the standard deviation characterizes the typical departure from the mean value.
 
Let us firstly, for simplicity, consider the estimators $\xi_{ML}^*$ and $\zeta^*_{ML}$. The expectation values for these estimators can be calculated in the following manner
\begin{eqnarray} 
\label{a11}
\begin{array}{l} 
\langle {\xi^*_{ML}} \rangle  = \langle {\left(\sum_{\ell}\sum_{Y}a_{\ell}^Y
d_{\ell}^Y\right)}/{\left(\sum_{\ell}\sum_{Y} a_{\ell}^{Y2}\right)}
\rangle = {\left(\sum_{\ell}\sum_{Y}a_{\ell}^Y
\langle d_{\ell}^Y \rangle \right)}/{\left(\sum_{\ell}\sum_{Y} a_{\ell}^{Y2}\right)} =1, \\
\langle {\zeta^*_{ML}} \rangle  =\langle {\left(\sum_{\ell}\sum_{Y}a_{\ell}^Y
d_{\ell}^Y b^*_{\ell}\right)}/{\left(\sum_{\ell}\sum_{Y}(a_{\ell}^Y b^*_{\ell})^2\right)}\rangle 
= {\left(\sum_{\ell}\sum_{Y}a_{\ell}^Y
\langle d_{\ell}^Y \rangle b^*_{\ell}\right)}/{\left(\sum_{\ell}\sum_{Y}(a_{\ell}^Y b^*_{\ell})^2\right)} = 0.
\end{array}
\end{eqnarray}
The angle brackets $\langle...\rangle$, in the above expression and elsewhere in the text, denote the ensemble average over the joint pdf (\ref{joint}). Furthermore, in this pdf, the input values for the tensor-to-scalar ratio and spectral index are chosen as $r=\hat{r}$ and $n_t=\hat{n}_t$ respectively. In deriving the above expressions we have firstly used (\ref{a10}) and (\ref{kk9}). We have also used the identity $\langle d_\ell^X \rangle = a_{\ell}^X$ which follows directly from  (\ref{c-sum}), (\ref{mean-variance}) and (\ref{a6}). Finally, in the bottom line, we have used the definition of the best pivot multipole (\ref{a9}). Similarly, the standard deviations can be calculated to yield
\begin{eqnarray} 
\label{a13}
\sigma_{{\xi^*_{ML}}}=
\xi_s^*, ~~~\sigma_{{\zeta^*_{ML}}}=
\zeta_s^*.
\end{eqnarray}

Proceeding in an identical manner, the expectation values and standard deviations for the maximum likelihood estimators $r^*_{ML}$, ${n_t}_{ML}$ and $\ln r_{ML}$ are given by
\begin{eqnarray}
\label{meanoverline}
\label{i9}
\begin{array}{l}
\langle {r^*_{ML}}\rangle =\hat{r}^*\langle {\xi^*_{ML}}\rangle = \hat{r}^*, 
\\
\langle {n_{t}}_{ML} \rangle = \hat{n}_{t}+\langle{\zeta^*_{ML}}\rangle = \hat{n}_t,
\\
\langle {\ln r_{ML}} \rangle =\ln \hat{r}^*+(\hat{n}_{t}-n_s+1)\ln(k_0/k_t^*)=\ln \hat{r},
\end{array}
\end{eqnarray}
and
\begin{eqnarray}
\label{sigma-r-nt}
\begin{array}{l}
\sigma_{{r^*_{ML}}}=\hat{r}^*\sigma_{{\xi^*_{ML}}}=\hat{r}^*{\xi^*_{s}},\\
\sigma_{{n_{t}}_{ML}}=\sigma_{{\zeta^*_{ML}}}={\zeta^*_{s}}, \\
\sigma_{{\ln r_{ML}}}\simeq \sqrt{(\xi_s^*)^2+(\ln(k_0/k_t^*)\zeta_s^*)^2}.
\end{array}
\end{eqnarray}
As expected, from expression (\ref{meanoverline}) it can be seen that the constructed $ML$ estimators are unbiased. Furthermore, the standard deviation of the estimator $\sigma_{{\ln r_{ML}}}$ strongly depends on the choice of the pivot multipole $k_0$, and is minimal for the choice $k_0=k_t^*$. We shall numerically verify these results in the following section.


\subsection{The dependence of results on cosmological parameters and experimental noises \label{s3.3.2}}

Let us now address the question of detection of RGWs in various CMB experiments. In order to quantify the ability to detect the signature of RGWs in the CMB data, it is convenient to define the signal-to-noise ratio as follows \cite{ours,ours2}
\begin{eqnarray}
\label{k12}
S/N\equiv \frac{\hat{r}^*}{\Delta{r^*}}.
\end{eqnarray}
Using expression (\ref{k11}) we arrive at an elegant expression for the signal-to-noise ratio
\begin{eqnarray}
\label{snr}
S/N=\sqrt{\sum_{\ell}\sum_{Y}\left(\frac{\hat{C}_{\ell,t}^Y}{\hat{\sigma}_{D_{\ell}^Y}}\right)^2}.
\end{eqnarray}
Thus, the signal-to-noise ratio contains contributions from individual power spectra and individual multipoles. These contributions have a clear physical meaning. For a particular power spectrum and a particular multipole, they represent the ratio of the expected signal due to RGWs to the overall uncertainty.

As was mentioned in Section \ref{s3.2}, for the analytical estimations, we had assumed $\xi^*_{p}\gg\xi^*_{s}$ (see (\ref{problem})). We can now relate this condition to the requirement on the value of the signal-to-noise ratio $S/N$.  Using Eqs.(\ref{a10}), (\ref{meanvaluernt}) and (\ref{k11}), we find that 
\begin{eqnarray}
\label{solve-problem}
\frac{\xi^*_{p}}{\xi^*_s}=\frac{r^*_{ML}}{\Delta r^*}\simeq \frac{\hat{r}^*}{\Delta r^*}=S/N.
\end{eqnarray}
Hence, the condition $\xi^*_{p}\gg\xi^*_{s}$ corresponds to the requirement $S/N\gg1$, i.e. to the requirement that the RGW signal may be well determined at a high signal-to-noise ratio.

In the discussion above we have mentioned that the best pivot multipole $\ell_t^*$, the signal-to-noise ratio $S/N$ and the uncertainty in determination of the RGW spectral index $\Delta n_t$ depend on the input cosmological model and the specifics of the CMB experiment. Let us analyze this dependence in more detail.

\begin{figure}[t]
\centerline{\includegraphics[width=20cm,height=7cm]{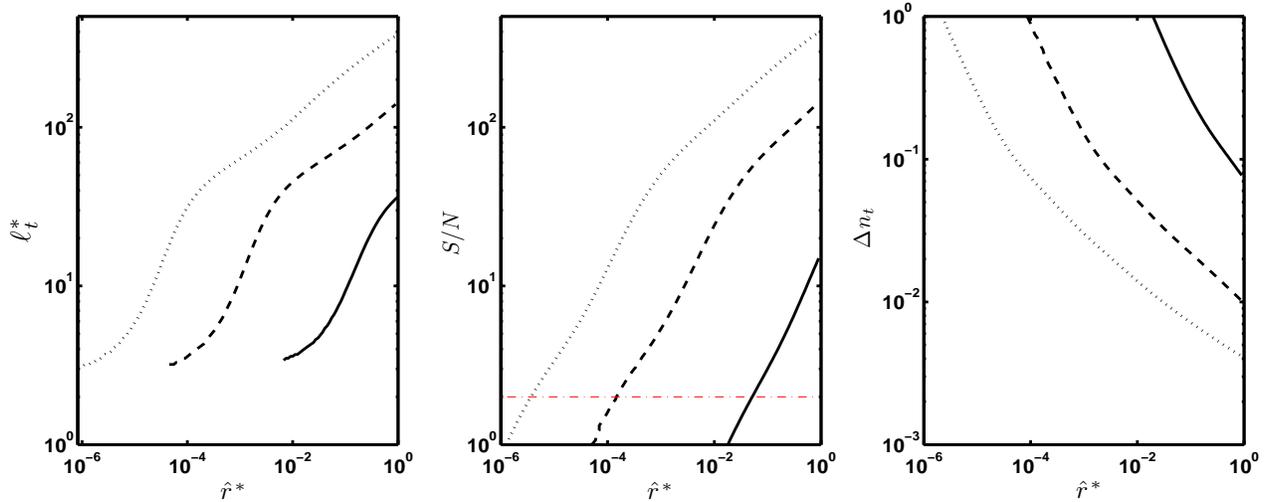}}
\caption{The figures show the value of the best pivot multipole
$\ell^*_{t}$ (left panel), signal-to-noise ratio $S/N$(middle panel) and
the uncertainty in the RGW spectral index $\Delta{n_t}$ (right panel) 
as functions of  the tensor-to-scalar ratio $\hat{r}^*$.
The solid lines correspond to the Planck instrumental noises (see (\ref{planck}));
the dashed lines correspond to noises from cosmic lensing (see (\ref{ideal1}));
and the dotted lines correspond to reduced cosmic lensing noise (see (\ref{ideal1})).
}\label{figure1}
\end{figure}

The input cosmological model is determined by specifying the background cosmological model, along with the parameters determining the density perturbations and gravitational waves. The background cosmological parameters and contribution from density perturbations are fairly well constrained by the current observations \cite{WMAP3}. The variation of these parameters within the margin allowed by these constraints will not significantly alter our results. For this reason, we shall fix the background cosmological model using the values of the typical $\Lambda$CDM model (\ref{background}). We shall also fix the contribution of density perturbations at a value (\ref{dp}). Furthermore, numerical calculations show that the dependence of various parameters on the input value of the spectral index $\hat{n}_t$ is weak, for this reason in evaluations of this section we shall set $\hat{n}_{t}=0$. Thus, we shall be interested in the dependence of the parameters on value of the input tensor-to-scalar ratio $\hat{r}$. FIG. \ref{figure1} shows the values of quantities $\ell_t^*$, $S/N$ and $\Delta{n_t}$ as functions of $\hat{r}^*$, calculated using the expressions (\ref{a9}), (\ref{snr}) and (\ref{k11}).

As was explained in Section \ref{s-spectra}, the specifics of the CMB experiment are determined by the noise power spectra, the cut sky factor and window function. In this section we shall consider the parameters $\ell_t^*$, $S/N$ and $\Delta n_t$ for the three cases specified in Section \ref{s-spectra} (see (\ref{planck}) and (\ref{ideal1})). The different curves (solid, dashed and dotted) on the three panels in FIG. \ref{figure1} show the corresponding values of quantities $\ell_t^*$, $S/N$ and $\Delta{n_t}$ for these three noise scenarios.


\begin{figure}[t]
\centerline{\includegraphics[width=14cm,height=10cm]{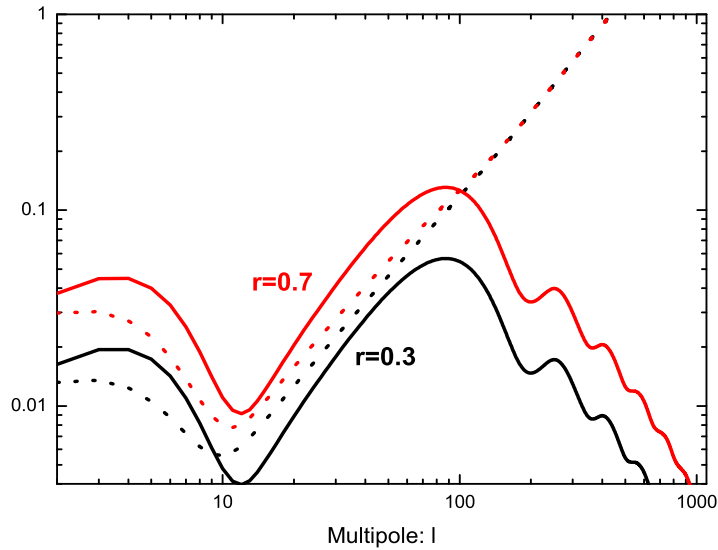}}
\caption{The comparison of $\hat{C}_{\ell}^{B}$ and $\hat{\sigma}_{D_{\ell}^{B}}$ (which enter the expression for signal-to-noise ratio $S/N$ (\ref{snr})), for models with $\hat{r}^*=0.3$ (black) and $\hat{r}^*=0.7$ (red). The solid lines show the `signal'-term (i.e.~power spectrum) $\ell(\ell+1)\hat{C}_{\ell}^B/2\pi(\mu$K$^2)$, and dotted lines show the `noise'-term $\ell(\ell+1)\hat{\sigma}_{D_{\ell}^B}/2\pi(\mu$K$^2)$. The quantity $\hat{\sigma}_{D_{\ell}^{B}}$ was calculated using the Planck noises (\ref{planck}).
}\label{figurea2}
\end{figure}

The left panel of FIG. \ref{figure1} shows the best pivot multipole $\ell_t^*$ as a function of the input tensor-to-scalar ratio $\hat{r}^*$ which is defined with respect to the best pivot multipole. It can be seen that, for small values of  $\hat{r}^*$, the best pivot multipole $\ell_t^*$ is small. This behaviour can be easily understood. For small values of $\hat{r}^*$, the constraints on $r^*$ and $n_t$ mainly come from $B$-mode power spectrum \cite{ours2}. However, in the $B$-mode  the main contribution to the signal comes from large angular scales corresponding to $\ell\lesssim10$, where the signal is mainly due to cosmic reionization \cite{ours,ours2}. Thus, for small $\hat{r}^*$, the constraints on parameters $r$ and $n_t$ are most stringently determined at large angular scales corresponding to multipoles $\ell\lesssim 10$. For this reason, for small values of $\hat{r}^*$ the best pivot multipole is small, corresponding to the scale at which the parameters $r$ and $n_t$ are most stringently determined. On the other hand, for large values $\hat{r}^*$, the best pivot multipole $\ell_t^*$ also becomes large. This happens due to two reasons. Firstly, with an increase in value of $\hat{r}^*$, the relative contribution of the reionization contribution to the to the $S/N$ decreases, while the relative contribution of the multipoles around $\ell\approx 90$ (where the $B$-mode spectrum is expected to have a maximum) increases (see FIG. \ref{figurea2}). Thus, the contribution of RGWs at higher multipoles ($\ell\sim100$) becomes significant, which in turn increases the value of the best pivot multipole. Secondly, when $\hat{r}^*$ is large, the contributions from the $C,T,E$ power spectra become important in constraining $r$ and $n_t$ \cite{ours2}. For these power spectra, the main contribution to the signal comes from the multipoles $10\lesssim\ell\lesssim100$ \cite{ours2}. This again leads to an increase in the value of $\ell_t^*$.

The middle panel in FIG. \ref{figure1} shows the signal-to-noise ratio $S/N$ as a function of $\hat{r}^*$. As expected, the signal-to-noise ratio rises with the increase of $\hat{r}^*$. Setting the threshold value of $S/N=2$, we can determine the detection possibilities for the three considered examples: $\hat{r}^* \geq 0.05$ for Planck noises; $\hat{r}^* \geq 1.5\times 10^{-4}$ for the case with cosmic lensing; $\hat{r}^* \geq 3.7\times 10^{-6}$ for the case with the  reduced cosmic lensing.
These estimations are consistent with previous results \cite{lensing1,lensing2,wuran4,ours}.

Finally, the right panel in FIG. \ref{figure1} presents the achievable constraints on the spectral index $\Delta n_t$ as a function of $\hat{r}^*$. As expected, the uncertainty in determining the spectral index drops with the increase of the input value $\hat{r}^*$. For the case of Planck mission, the uncertainty in estimation of $n_t$ always remains fairly large. Even for large value $\hat{r}^*=1$ the constraint on the spectral index is $\Delta n_t=0.08$ (for comparison, the Planck mission will be able to achieve constraint of $\Delta n_s=0.0045$ on the spectral index of density perturbations \cite{Planck}). For a value $\hat{r}^*=0.1$, the constraint on the spectral index is $\Delta n_t=0.25$, which is too large to constrain inflationary models or to verify the consistency relation. Potentially, in an idealized situation with reduced cosmic lensing, for $\hat{r}^*=0.1$, we can constrain the spectral index to the level $\Delta n_t=0.007$. If this accuracy can be achieved in the future, it will place a fairly tight constraint on inflationary models.


\section{Comparison with numerical simulations\label{s4}}

In Section \ref{s3} we have analytically studied the likelihood analysis of the RGW parameters $r^*$, $n_t$ and $r$, as well as introduced the best pivot multipole $\ell_t^*$ (corresponding to the best pivot wavenumber $k_t^*$) and explained its significance. We have analytically derived expressions for the uncertainties of the RGW parameters and the value of the best pivot multipole, in terms of the CMB power spectra, experimental uncertainties and the estimators of the CMB power spectra. In this section we shall compare the analytical results of the previous section with numerical simulations. We shall show that, although we have used a number of approximations, the analytical results are in good agreement with the exact numerical results based on the analysis of simulated data.

This section is separated into two parts. In the first subsection, using a single simulated data set $\{D_{\ell}^X\}$, we shall use the Markov Chain Monte Carlo (MCMC) techniques to construct the posterior pdf for the RGW parameters. We shall calculate the uncertainties and correlations associated with the parameters, and compare these values with the analytical predictions in Sections \ref{s3.3} and \ref{s3.4}. In the second subsection we shall generate $300$ samples of data sets $\{D_{\ell}^X\}$. For each individual sample, using the posterior pdf $P(r^*,n_t)$ we shall calculate the estimates for the RGW parameters ${r_{ML}^*}$, ${n^*_{tML}}$ and ${r}_{ML}$. Analyzing the distribution of these estimates, we shall evaluate the mean values and the standard deviation, and compare these with the analytical predictions from Section \ref{section-manyrealizations}.


\subsection{Likelihood analysis of a single simulated data set\label{s4.1}}

In this subsection, from a single simulated data set $\{D_{\ell}^X\}$, using the likelihood analysis procedure, we shall derive the constraints on the tensor-to-scalar ratio and the RGWs primordial spectral index.

\begin{figure}[t]
\centerline{\includegraphics[width=18cm,height=12cm]{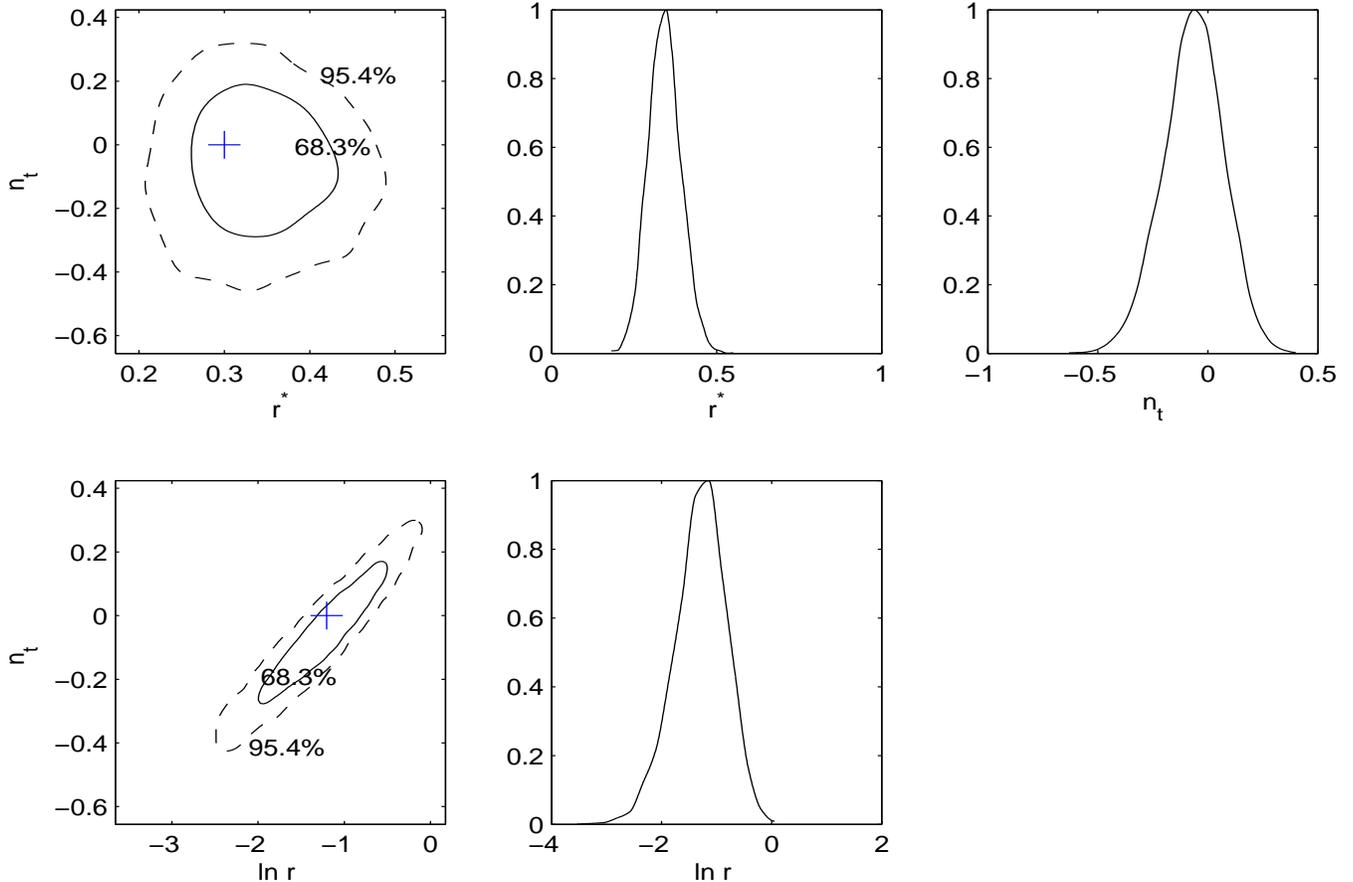}}
\caption{2-dimension and 1-dimension posterior constraints for parameters: $r^*$ and $n_t$ (upper panels), and for parameters: $\ln r$ and $n_t$ (lower panels).
The blue $'+'$ in the left panels indicate the value of the input model parameters.}\label{figure3}
\end{figure}

In order to simulate the CMB data, we shall randomly draw a data set $\{D_{\ell}^X\}$, from an underlying pdf (\ref{joint}) (see Appendix \ref{appendix-a}). This pdf depends on the input cosmological model and characteristics of the CMB experiment. We shall choose as an input cosmological model, a model with the background cosmological parameters given in (\ref{background}) and the contribution of density perturbation (\ref{dp}). The input parameters for the RGW field will be chosen as
\begin{eqnarray}
\label{gw}
\hat{r}=0.3,~~\hat{n}_{t}=0.0. 
\end{eqnarray}
To characterize the properties of the CMB experiment, namely the power spectra of noises, the cut sky factor and the window function, we shall adopt the values specified for the Planck satellite mission (\ref{planck}) \cite{Planck}.

In order to simulate and analyze the data, we proceed as follows:\\
1) We generate a single data sample $\{D_{\ell}^Y|Y=C,T,E,B;\ell=2,3,\cdot\cdot\cdot,1000\}$, drawn from the pdf (\ref{joint}). \\
2) Using (\ref{a9}), we calculate the best pivot multipole scale $\ell_t^*=21.1$ (corresponding to the best pivot wavenumber $k_t^*=0.002$Mpc$^{-1}$). Note that, the value of $\ell_t^*$ does not depend on the concrete realization generated in Step 1. \\
3) Using the MCMC method (see \cite{cosmomc} for details), we construct the likelihood function $\mathcal{L}$ as a function of two free parameters $r^*$ and $n_t$, with the other cosmological parameters fixed at their ``best-fit" values given by (\ref{background}) and (\ref{dp}). Choosing a uniform prior we build the posterior pdf  $P(r^*,n_t)$ (which is exactly equal to the likelihood function $\mathcal{L}$ (see (\ref{k3}) and (\ref{k4}))). \\
4) Using the posterior pdf  $P(r^*,n_t)$, we find the maximum likelihood values $({r}^*_{ML},{n_t}_{ML})$, and plot the contours corresponding to $68.3\%$ and $95.4\%$ confidence interval regions in the $(r^*,n_t)$ plane surrounding these values. We also calculate the 1-dimensional posterior pdfs for variables $r^*$ and $n_t$. From  $P(r^*,n_t)$, we calculate the uncertainties $\Delta r^*$ and $\Delta n_t$. Using the importance sample technique (see \cite{cosmomc,bayesian}), we evaluate the correlation coefficient  $\rho_{(r^*,n_t)}$ defined in (\ref{k14}).\\
5) We now choose a different value of the pivot wavenumber $k_0=0.05$Mpc$^{-1}$, corresponding to the value for the pivot multipole $\ell_0=500$. Using (\ref{j3}), we calculate the tensor-to-scalar ratio for this pivot wavenumber $r$ as a function of the parameters $r^*$ and $n_t$. From the posterior probability function $P(r^*,n_t)$, using the importance sample technique, we can obtain the uncertainty $\Delta \ln r$ and the correlation coefficient $\rho_{(n_t,\ln r)}$ defined in (\ref{lnr-rho}).

The results of the simulation and analysis is shown in FIG. \ref{figure3}. The panels on the top show the constraints in the $r^*-n_t$ plane (top-left), and the 1-dimensional posterior pdfs for $r^*$ (top-middle) and $n_t$ (top-right). The constraint on the parameters $r^*$ and $n_t$, together with the correlation coefficient are as follows
\begin{eqnarray}
\label{simulation1}
\begin{array}{c}
r^*=0.343^{+0.047}_{-0.053},~(68.3\% {\rm C.L.});~~{n_{t}}=-0.067^{+0.146}_{-0.130},~(68.3\% {\rm C.L.});~~\rho_{(r^*,n_t)}=-0.02. 
\end{array}
~~({\rm simulation~results})
\end{eqnarray}
For comparison, the analytical formulae (\ref{meanvaluernt}), (\ref{k11}) and (\ref{k14}) yield the following results for these quantities
\begin{eqnarray}
\label{analytic1}
\begin{array}{c}
{r^*_{ML}} \pm\Delta r^*=0.345\pm0.047;~~{{n_{t}}_{ML} \pm \Delta n_t}=-0.062\pm0.135;~~ \rho_{(r^*,n_t)} =0. \end{array}
~~({\rm analytical~results})
\end{eqnarray}
As can be seen, the analytical results (\ref{analytic1}) are in good agreement with results of simulation (\ref{simulation1}).

The bottom panels in FIG. \ref{figure3} show the constraints in the $\ln r-n_t$ plane (bottom-left), and the 1-dimensional posterior pdfs for $\ln r$ (bottom-middle). As expected, the confidence interval contours in the $\ln r-n_t$ indicate a strong correlation between $\ln r$ and $n_t$. The corresponding constraints and correlation coefficient are as follows
\begin{eqnarray}
\label{simulation2} 
\begin{array}{c}
\ln r=-1.299^{+0.527}_{-0.413}, ~~(68.3\% {\rm C.L.});
~~
\rho_{(n_t,\ln r)}=0.95.~~({\rm simulation~results}) 
\end{array}
\end{eqnarray}
The analytical expressions (\ref{overliner}), (\ref{j5}) and (\ref{ff2}), yield the following results for these quantities 
 \begin{eqnarray}\label{analytic2}\begin{array}{c}
 {\ln r_{ML}\pm\Delta \ln r}=-1.307\pm0.456;~~ \rho_{(n_t,\ln r)} =0.94.~~({\rm analytical~results})
  \end{array}
 \end{eqnarray}
 Once again, we find that analytical and the exact results are consistent with each other.

Furthermore, we have applied the same simulation and analysis procedure to the case with the ``cosmic lensing" noises (see (\ref{ideal1})), for input values of tensor-to-scalar ratio $\hat{r}^*=0.1$, $0.2$ and $0.3$.  We found that in all these cases, the numerical estimations for $\Delta {r^*}$, $\Delta{\ln r}$ and $\Delta{n_t}$ agree with the analytical expression to within $20\%$.
Thus the analytical formulae for $\Delta {r^*}$, $\Delta{\ln r}$ and $\Delta{n_t}$ seem to be accurate.


\subsection{Maximum likelihood analysis in numerous data simulations\label{s4.2}}

In this subsection, we shall discuss the distribution of the maximum likelihood estimators for the RGW parameters ${r}^*_{ML}$, ${r}_{ML}$ and ${n_t}_{ML}$ in multiple realizations. We shall generate a simulated CMB data set $\{D_{\ell}^X\}$ a number of times. For each individual realization we shall calculate the estimators ${r}^*_{ML}$, ${r}_{ML}$ and ${n_t}_{ML}$. We shall then analyze the distribution of these parameters and compare these results with analytical calculations.

\begin{figure}[t]
\centerline{\includegraphics[width=18cm,height=10cm]{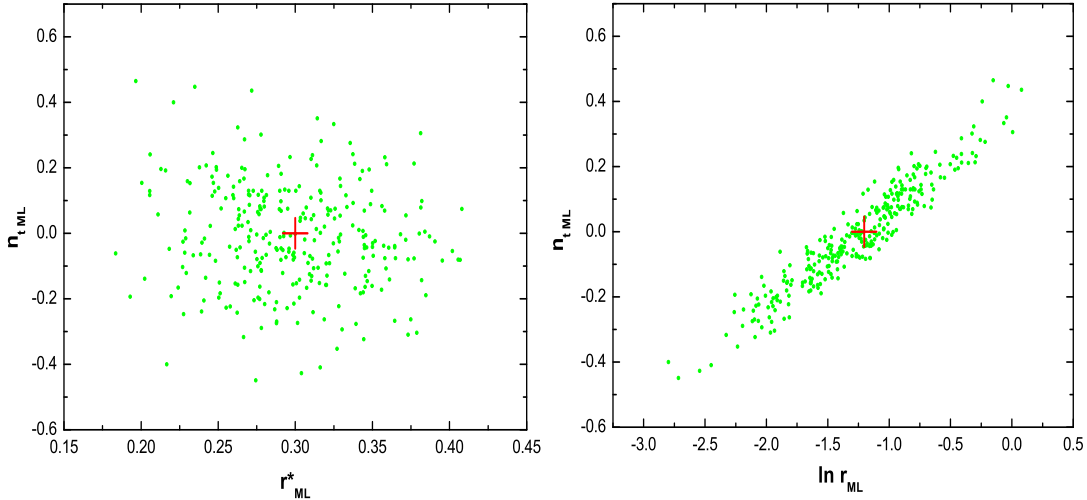}}
\caption{The values of the ML estimators from 300 simulations are shown projected onto the ${n_{t}}_{ML}-r^*_{ML}$ plane (left panel), and ${n_{t}}_{ML}-\ln r_{ML}$ plane (right panel). The red $'+'$ indicate the value of the input model parameters.}\label{figure2}
\end{figure}

In order to generate and analyze the data, we proceed in the following manner:\\
1) A collection of 300 samples of data sets $\{D_{\ell}^Y|Y=T,E,B,C;\ell=2,3,\cdot\cdot\cdot,1000\}$ is randomly generated from an underlying pdf $f(D_{\ell}^{C},D_{\ell}^{T},D_{\ell}^{E},D_{\ell}^{B})$, given in (\ref{joint}). The input cosmological model and the noise characteristics of the CMB experiment are chosen in the same manner as in Section \ref{s4.1}. \\
2) Using (\ref{a9}),  we calculate the best pivot multipole $\ell_t^*=21.1$ (corresponding to the best pivot wavenumber $k^*_t=0.002$Mpc$^{-1}$). Note that, the value of $\ell_t^*$ does not depend on the concrete realization generated in Step 1.\\
3) For each individual sample, we construct the likelihood function $\mathcal{L}$ as a function of variables $r^*$ and $n_t$, which is equal to the posterior pdf $P(r^*,n_t)$ (see (\ref{k3}) and (\ref{k4}))). For each individual sample, an automated search (which uses the numerical technique of the simulated annealing \cite{numerical}) determines the maximum likelihood estimators ${r^*_{ML}}$ and ${n_t}_{ML}$ (at which the posterior pdf $P(r^*,n_t)$ reaches a maximum). The calculated values ${r^*_{ML}}$ and ${n_t}_{ML}$ are plotted in FIG. \ref{figure2} (left panel).\\
4) We adopt a different pivot wavenumber $k_0=0.05$Mpc$^{-1}$, corresponding to the value for the pivot multipole $\ell_0=500$. From the set of values $(r^*_{ML},{n_{t}}_{ML})$, we calculate the corresponding values of tensor-to-scalar ratio for the new pivot wavenumber $r_{ML}$ using (\ref{j3}). The resulting values are illustrated in FIG. \ref{figure2} (right panel).

The mean values and standard deviations for the quantities ${r^*_{ML}}$ and ${n_t}_{ML}$ (shown in FIG. \ref{figure2} (left panel)), obtained from the analysis of the simulated data, are
\begin{eqnarray}
\label{simulation3}
\langle{r^*_{ML}}\rangle \pm\sigma_{r^*_{ML}}=0.298\pm0.046;~~\langle {{n_{t}}_{ML}}\rangle \pm\sigma_{{n_{t}}_{ML}}=-0.001\pm0.152.~~({\rm simulation~results})
\end{eqnarray}
For comparison, we can calculate the corresponding quantities using the analytical expressions derived in Section \ref{s3}.  Using (\ref{meanoverline}) and (\ref{sigma-r-nt}), we obtain 
 \begin{eqnarray}\label{analytic3}
\langle{r^*_{ML}}\rangle \pm\sigma_{r^*_{ML}}=0.300\pm0.047;~~ 
\langle {{n_{t}}_{ML}}\rangle 
\pm\sigma_{n_{tML}}=0\pm0.135.~~({\rm analytical~results})
 \end{eqnarray}
Comparing (\ref{analytic3}) with (\ref{simulation3}), we find that the analytical expressions are in good agreement with results of numerical simulation.

In a similar fashion, for the mean values and the standard deviation of quantity ${r}_{ML}$ (shown in FIG. \ref{figure2} (right panel)), we obtain  
\begin{eqnarray}
\label{simulation4}
\langle{\ln r_{ML}}\rangle \pm \sigma_{\ln r_{ML}}= -1.252\pm0.508.~~({\rm simulation~result})
\end{eqnarray}
The analytical expressions (\ref{i9}) and (\ref{sigma-r-nt}) yield the following results
\begin{eqnarray}
\label{analytic4}
\langle{\ln r_{ML}}\rangle \pm \sigma_{\ln r_{ML}}=-1.204\pm0.456.~~({\rm analytical~result})
\end{eqnarray}
Comparing (\ref{simulation4}) with (\ref{analytic4}), we find a reasonable agreement to within $10\%$.


\section{Conclusion\label{s5}}

In this paper, we have analyzed the potential joint constraints on the two parameters characterizing the RGW background, the tensor-to-scalar ratio $r$ and the tensor primordial spectral index $n_t$, achievable by the upcoming CMB observations. We have shown that, in general, there exists a correlation between the parameters $r$ and $n_t$.  However, when considering the tensor-to-scalar ratio $r^*$ defined at the best pivot multipole number $\ell_t^*$, the correlation between $r^*$ and $n_t$ disappears. Furthermore, the uncertainty $\Delta r^*$ has the least possible value. We have derived analytical formulae for calculating $\ell_t^*$, $\Delta r^*$, $\Delta n_t$, $\Delta r$, and the correlation coefficient between $r$ and $n_t$. Using numerical simulations of future CMB data we have verified the robustness of our analytical estimations and have shown that our fairly simple analytical expressions agree with exact numerical evaluations to within $20\%$.
We have also discussed the dependence of our results on the background cosmological model, the amplitude of the RGWs, and the characteristics of the CMB experiment. We have studied the dependence of the signal-to-noise ratio $S/N$ along with the value of the best pivot multipole $\ell_t^*$ and the uncertainty $\Delta n_t$ on the amplitude of the RGWs. We show that, although the Planck satellite will potentially be able to measure the tensor-to-scalar ratio to a level $r\gtrsim0.05$ (at $2\sigma$ C.L.), the uncertainty in determining the spectral index will remain fairly large $\Delta n_t \gtrsim 0.25$ (for $r=0.1$). Thus, for example, the Planck satellite will not be able to verify the so-called consistency relation $n_t=-r/8$. In an idealized scenario, where the noises are limited by reduced cosmic lensing noise, the precision $\Delta n_t \gtrsim 0.007$ (for $r=0.1$) is achievable, thus, potentially allowing tight constraints on possible inflationary scenarios. The analytical results presented here provide a simple and quick method to investigate the ability of the future CMB observations to detect RGWs.

~

{\bf Acknowledgement:}
The authors thank L.~P.~Grishchuk for helpful discussions and useful suggestions. W.~Zhao is partly supported by Chinese NSF under grant Nos.~10703005 and 10775119. In this paper, we have used the CAMB code for calculating the various CMB power spectra \cite{camb}.


\appendix

\section {Exact probability density functions for $D_{\ell}^{Y}$ and likelihood function \label{appendix-a}}

In \cite{ours} (see also \cite{ours2,wishart2,pdf1}) we have derived the pdfs for the best unbiased estimators $D_{\ell}^{Y}$ of the various CMB power spectra $C_{\ell}^{Y}$. These were derived under the assumption that the primordial perturbations (density perturbations and RGWs) are isotropic and homogeneous Gaussian random fields, and that the noises associated with the CMB measurements can be assumed Gaussian. In this appendix we shall briefly list the main results that have been used in the present paper.

The joint pdf for the estimators $D_{\ell}^{T}$, $D_{\ell}^E$, $D_{\ell}^{B}$ and $D_{\ell}^C$ has the following form
\begin{eqnarray}
\label{joint}
f(D_{\ell}^{C},D_{\ell}^{T},D_{\ell}^{E},D_{\ell}^{B})=
f(D_{\ell}^{C},D_{\ell}^{T},D_{\ell}^{E})f(D_{\ell}^{B}),
\end{eqnarray}
where the pdf $f(D_{\ell}^{B})$ has the form of the $\chi^2$ distribution,
\begin{eqnarray}
\label{chi2}
f(D_{\ell}^{B})=\frac{(n_{\rm e}W^2_{\ell})v^{(n_{\rm e}-2)/2}e^{-v/2}}{2^{n_{\rm e}/2}\Gamma(n_{\rm e}/2)(\sigma_\ell^B)^2},
\end{eqnarray}
and the joint pdf $ f(D_{\ell}^{C},D_{\ell}^{T},D_{\ell}^{E})$ is the Wishart distribution
\begin{eqnarray}
\label{wishart}
\begin{array}{c}
f(D_{\ell}^{C},D_{\ell}^{T},D_{\ell}^{E})=\left\{\frac{1}{4(1-{\rho_{\ell}^2})({\sigma_\ell^T}{\sigma_\ell^E})^2}\right\}^{n_{\rm e}/2}
\frac{ (n_{\rm e}W_{\ell}^2)^3({x}{y}-{z}^2)^{(n_{\rm e}-3)/2}}{\pi^{1/2}\Gamma(n_{\rm e}/2)\Gamma((n_{\rm e}-1)/2)} \\ \times
\exp\left\{-\frac{1}{2(1-{\rho_\ell^2})}\left(\frac{{x}}{(\sigma_\ell^T)^2}
+\frac{{y}}{(\sigma_\ell^E)^2}-\frac{2{\rho_l} {z}}{{\sigma_\ell^T}{\sigma_\ell^E}}\right)\right\}.
\end{array}
\end{eqnarray}
In the above expressions (\ref{chi2}) and (\ref{wishart}), $C_{\ell}^{Y}$ are the corresponding CMB power spectra, $N_{\ell}^{Y}$ are the noise power spectra, and $W_{\ell}$ is the window function. The quantity $n_{\rm e} = (2\ell+1)f_{\rm sky}$ is the effective degree of freedom for a particular multipole $\ell$ in the case of partial sky coverage with the cut sky factor $f_{\rm sky}$. The quantities $v$, $x$, $y$, $z$ are defined as follows
\begin{eqnarray}
\nonumber
\begin{array}{l}
 v\equiv n_{\rm e}(D_{\ell}^{B}W_{\ell}^2+N_{\ell}^{B})/(C_{\ell}^{B}W_{\ell}^2+N_{\ell}^{B}),\\
{x}\equiv n_{\rm e}(D_\ell^{T}W_{\ell}^2+N_{\ell}^{T}), ~~
{y}\equiv n_{\rm e}(D_\ell^{E}W_{\ell}^2+N_{\ell}^{E}),~~
{z}\equiv n_{\rm e}D_\ell^{C}W_{\ell}^2.
\end{array}
\end{eqnarray}
In (\ref{chi2}), $\sigma_\ell^B$ is the standard deviation for the multipole coefficient $a_{\ell m}^B$. The quantities ${\sigma_\ell^T}$, ${\sigma_\ell^E}$ and $\rho_{\ell}$ in (\ref{wishart}) are correspondingly the standard deviations and the correlation coefficient for the multipole coefficients $a_{\ell m}^{T}$ and $a_{\ell m}^{E}$. These are expressible in terms of the CMB and noise power spectra in the following form
\begin{eqnarray}
\nonumber
\begin{array}{c}
{\sigma_\ell^T}= \sqrt{ C_\ell^{T}W_{\ell}^2+N_\ell^{T}},~~
\sigma_\ell^E=\sqrt{C_\ell^{E}W_{\ell}^2+N_{\ell}^{E}},~~
\sigma_\ell^B=\sqrt{C_\ell^{B}W_{\ell}^2+N_{\ell}^{B}},\\
{\rho_\ell}=C_\ell^{C}/\sqrt{(C_\ell^{T}+N_{\ell}^{T}W_{\ell}^{-2})(C_\ell^{E}+N_{\ell}^{E}W_{\ell}^{-2})}.
\end{array}
\end{eqnarray}
Finally, the likelihood function $\mathcal{L}$ introduced in Section \ref{s3.1} is, up to a constant of normalization, the product of the joint pdf $f(D_{\ell}^{C},D_{\ell}^{T},D_{\ell}^{E},D_{\ell}^{B})$, i.e.
\begin{eqnarray}\label{exact-likelihood}
\mathcal{L}\propto \prod_{\ell} f(D_{\ell}^{C},D_{\ell}^{T},D_{\ell}^{E},D_{\ell}^{B}).
\end{eqnarray}


\end{document}